\documentclass[prd,twocolumn,showpacs,amsmath,amssymb]{revtex4}

\usepackage{graphicx}

\begin{document}

\title{\boldmath
Review of Experimental Studies of $\psi(3770)$ non-$D\bar D$ Decays
}
\author{ G. Rong,~~D. Zhang,~~J.C. Chen \\
Institute of High Energy Physics, Beijing 100049, People's Republic of China}

\begin{abstract}
We review the progress on experimental studies of the
non-$D\bar D$ decays of the $\psi(3770)$ resonance. 
With the world average of the observed cross sections for
$D\bar D$ production measured at 3.773 GeV by the MARK-I, MARK-II, BES and CLEO
Collaborations, combined together with the cross section for $\psi(3770)$ production at its peak 
as well as initial state radiative correction factor,
we find that the non-$D\bar D$ branching fraction of $\psi(3770)$ decays is
$B[\psi(3770)\rightarrow {\rm non}-D\bar D]=(19.8\pm 1.8 \pm 5.6)\%$,
which is consistent within error with $B[\psi(3770)\rightarrow {\rm non}-D\bar D]=(14.7\pm 3.2)\%$
measured previously by the BES Collaboration. In addition,
a global amplitude analysis of the cross sections for $e^+e^- \rightarrow {\rm LH}$ (LH= light hadron)
measured by the CLEO Collaboration shows that the light hadron branching fraction of
$\psi(3770)$ decays can be as large as about $11\%$.
Combing the totally measured hadronic and electromegnatic transition rate together with the
light hadron branching fraction in the decays of $\psi(3770)$ yields the total non-$D\bar D$
branching fraction in the decays of $\psi(3770)$ to be about $13\%$.

\end{abstract}


\maketitle

\section{Introduction}

    The structure produced in $e^+e^-$ annihilation near 3.770 GeV 
found by the MARK-I Collaboration~\cite{I_Peruzzi_PRL39_1977_p1301_MARK-I} 
has been popularly interpreted as a charmonium 
with a dominated $1^3D_1$ wave of $c \bar c$ bound state with a small admixture of $2^3S_1$ wave.
This structure has been named as the $\psi(3770)$ resonance.
However, other explanations for this structure, such as its being a p-wave
four quark state~\cite{I_Peruzzi_PRL39_1977_p1301_MARK-I, four_quark_state} 
or a molecular $D\bar D$ threshold resonance are also conceivable.
Historically, the discovery of this structure by the MARK-I Collaboration and
the measurement of the resonance parameters of $\psi(3770)$ from the MARK-II
experiment~\cite{markii_psi3770_prd21_p2716} never rule out 
the molecular interpretation of this structure.
To better understand the nature of the $\psi(3770)$ resonance, we need to search for the
non-$D\bar D$ decay channels of the $\psi(3770)$ resonance.

If the $\psi(3770)$ resonance is really a pure $c\bar c$ bound state, 
the potential mode expects that 
more than $99\%$ of $\psi(3770)$ decay
into $D\bar D$ final states~\cite{E_Eichten_PRL34_1975_p369,E_Eichten_PRD21_1980_p203},
the non releativistic QCD (NRQCD) calculation predicts that the branching fraction for 
$\psi(3770)\rightarrow$ non-$D\bar D$ decays should be about $4\%$~\cite{GT_Chao_NRQCD_cal},
and the NRQCD+FSI (Final State Interaction) calculations predict the branching fraction for
$\psi(3770)\rightarrow$ non-$D\bar D$ decays should be within the range from $5.5\%$ to
$6.4\%$~\cite{Liu_Zhang_Li_PLB675_2009_p441}. 
However, if the $\psi(3770)$ resonance
contains four-quark admixture, the branching fraction 
for $\psi(3770)\rightarrow$ {non}-$D\bar D$ decays 
would be around $10\%$~\cite{BM_Voloshin_paper_prd71_2005_p114003}. 
In addition, if there are some new structure(s) around 3.773 GeV 
in addition to the dominated $1^3D_1$ wave of $c \bar c$ bound state, $\psi(3770)$
along there, the experimentally measured non-$D \bar D$ decay branching
fraction would also increase in the case of assuming that there is only one simple $\psi(3770)$ in 
the energy region between 3.70 and 3.80 GeV. 

Some models~\cite{four_quark_state} predict that there are exsiting
p-wave four quark state or molecular $D\bar D$ threshold resonance 
in the open-charm energy region. 
In experiment, if such p-wave four quark state or molecular $D\bar D$ threshold resonance
exsiting near the dominated $1^3D_1$ wave of $c \bar c$ state,
one may observe some unexpected properties on the $\psi(3770)$ production and decays.
So, experimental measurements of the branching fraction
for $\psi(3770)\rightarrow$ non-$D\bar D$ decays would provide some important information 
about the nature of the $\psi(3770)$ resonance and about 
whether there are some new structure around 3.773 GeV.

\section{Hadronic and E-M transitions}

\subsection{Hadronic transition}

\begin{figure}
\includegraphics[width=8.5cm,height=7.5cm]
{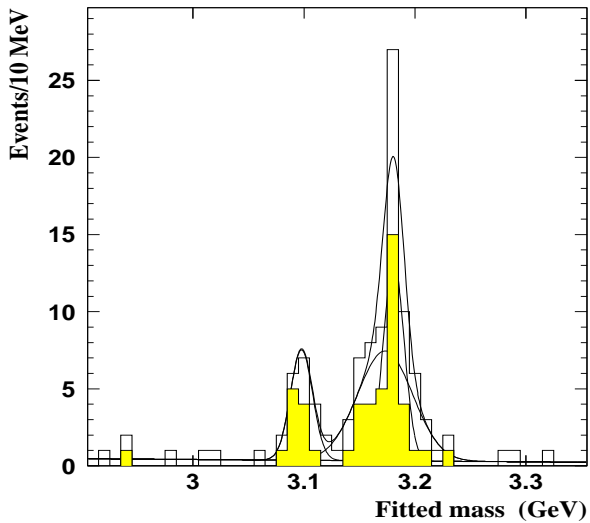}
\caption{The distribution of the dilepton masses for the
events of $l^+l^-\pi^+\pi^-$ from the data taken around 3.773 GeV,
the hatched histogram is for $\mu^+\mu^-\pi^+\pi^-$,
while the open one is for $e^+e^-\pi^+\pi^-$; the curves give the best fit
to the data; the first peak shows the events mainly coming from 
$\psi(3770)\rightarrow J/\psi \pi^+\pi^-$ while the second from
$\psi(3686)\rightarrow J/\psi \pi^+\pi^-$~\cite{PLB605_2005_63_BES}.
}
\label{psi3770_to_jpsipipi_bes}
\end{figure}

In 2003, the BES Collaboration
claimed the first observation of the first non-$D\bar D$ decay event
of $\psi(3770)$~\cite{PLB605_2005_63_BES},
that is $\psi(3770)\rightarrow J/\psi \pi^+\pi^-$, 
started experimentally studying
the non-open charm decays of the particles lying above the $D\bar D$ threshold.
From about 27.7 pb$^{-1}$ data taken near 3.773 GeV, the BES found $15 \pm 6$ events for
$\psi(3770)\rightarrow J/\psi \pi^+\pi^-$ non-$D\bar D$ 
decays as shown in figure~\ref{psi3770_to_jpsipipi_bes}, 
and measured the decay branching
$B[\psi(3770) \rightarrow J/\psi \pi^+\pi^-]=(0.34 \pm 0.14 \pm 0.09)\%$ corresponding to the
partial decay width of $\Gamma[\psi(3770) \rightarrow J/\psi \pi^+\pi^-]=(80\pm 33 \pm 23)$ keV. 

In 2005, the CLEO confirmed the BES measurement for this hadronic transition.
The CLEO measured the decay branching
$B[\psi(3770) \rightarrow J/\psi \pi^+\pi^-]=(0.189 \pm 0.20 \pm 0.20)\%$~\cite{PRL96_2006_082004_CLEO}.
Combining these two measurements, the PDG (Particle Data Group) 
gives the branching fraction of
$B[\psi(3770) \rightarrow J/\psi \pi^+\pi^-]=(0.193 \pm 0.28)\%$.
Later on, the CLEO found the $\pi^0\pi^0$ and
$\eta$ hadronic transitions of $\psi(3770)$.
The CLEO measurement of the branching fraction for $\pi^0\pi^0$ transition is 
$B[\psi(3770)\rightarrow J/\psi \pi^0\pi^0]=(8.0\pm 2.5\pm 1.6)\times 10^{-4}$~\cite{PRL96_2006_082004_CLEO} 
and branching fraction for $\eta$ transition is 
$B[\psi(3770)\rightarrow J/\psi \eta]=(8.7\pm 3.3\pm 2.2)\times 10^{-4}$~\cite{PRL96_2006_082004_CLEO}.

\subsection{E-M transition}

\begin{figure}
\includegraphics[width=7.5cm,height=7.0cm]
{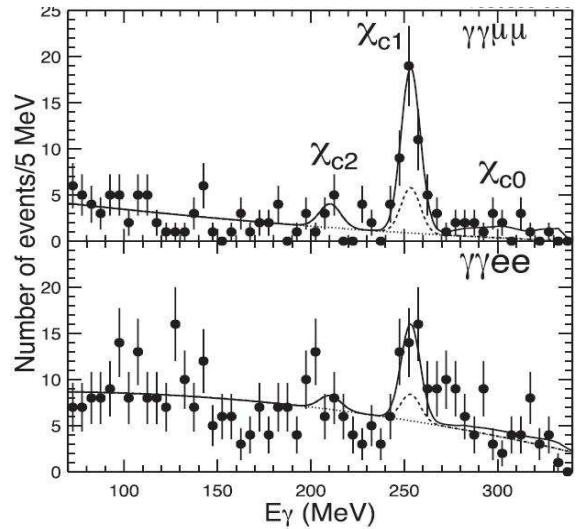}
\caption{Energy of the lower energy photon for the selected $e^+e^-\rightarrow \gamma\chi_{cJ}$,
where $\chi_{cJ} \rightarrow \gamma J/\psi$ while $J/\psi$ 
decay into $e^+e^-$ and $\mu^+\mu^-$~\cite{prl96_2006_p182001_cleo}.
}
\label{psi3770_to_gamma_chicj_cleo}
\end{figure}

In 2006, the CLEO found the electromagnetic transition signal events of 
$\psi(3770)\rightarrow \gamma\chi_{cJ}(J=0,1,2)$~\cite{prl96_2006_p182001_cleo}
by reconstructing the $\chi_{cJ}$ in the $\chi_{cJ}$ transition modes of $\gamma J/\psi$, 
where the $J/\psi$ decays to $e^+e^-$ or $\mu^+\mu^-$. 
Figure~\ref{psi3770_to_gamma_chicj_cleo} shows the energy of the photon from the transition
$\psi(3770)\rightarrow \gamma \chi_{cJ}$.
In addition to 
reconstruction of the $\chi_{cJ}$ in the transition modes,
the CLEO also reconstruct the $\chi_{cJ}$ in some hadronic final states of the $\chi_{cJ}$ decays.
The CLEO measurements of the branching fractions for $\psi(3770)\rightarrow \gamma\chi_{c0}$
and for $\psi(3770)\rightarrow \gamma\chi_{c1}$ are, respectively,
$B[\psi(3770)\rightarrow \gamma\chi_{c0}]=(0.73\pm 0.07\pm 0.06)\%$~\cite{prd74_2006_031106r_cleo} and
$B[\psi(3770)\rightarrow \gamma\chi_{c1}]=(0.28\pm 0.05\pm 0.04)\%$~\cite{prl96_2006_p182001_cleo}.
The CLEO set an upper limit of the branching fraction for
$\psi(3770)\rightarrow \gamma\chi_{c2}$ decay to be less than $0.9\%$ at $90\%$ C.L.. 

\section{Inclusive Decays}

\subsection{$B[\psi(3770)\rightarrow {\rm non-}D\bar D]$ measured at the BES-II}

    In assumption of that there is no additional structure around 3.773 GeV, 
and there is no unknown dynamics effects affecting $\psi(3770)$ production and decays,
the BES Collaboration studied the $\psi(3770)$ production and decays extensively.
By analyzing several different data samples taken at 3.650 GeV, 3.773 GeV and 
the data samples taken in the energy range from 3.650 to 3.872 GeV
with different analysis methods,
the BES Collaboration measured the branching fractions for $\psi(3770)\rightarrow$$D \bar D$ and
for $\psi(3770)\rightarrow$ non-$D \bar D$ for the first 
time~\cite{besii_psi3770_non_dd_1,besii_psi3770_non_dd_2,besii_psi3770_non_dd_3,besii_psi3770_non_dd_4}.
The averages of these branching fractions are
$$B[\psi(3770) \rightarrow D\bar D]=(85.3 \pm 3.2)\%,$$
\noindent
and
$$B[\psi(3770) \rightarrow {\rm non-}D\bar D]=(14.7 \pm 3.2)\%.$$
\noindent 

\begin{figure}
\includegraphics[width=9.0cm,height=8.0cm]
{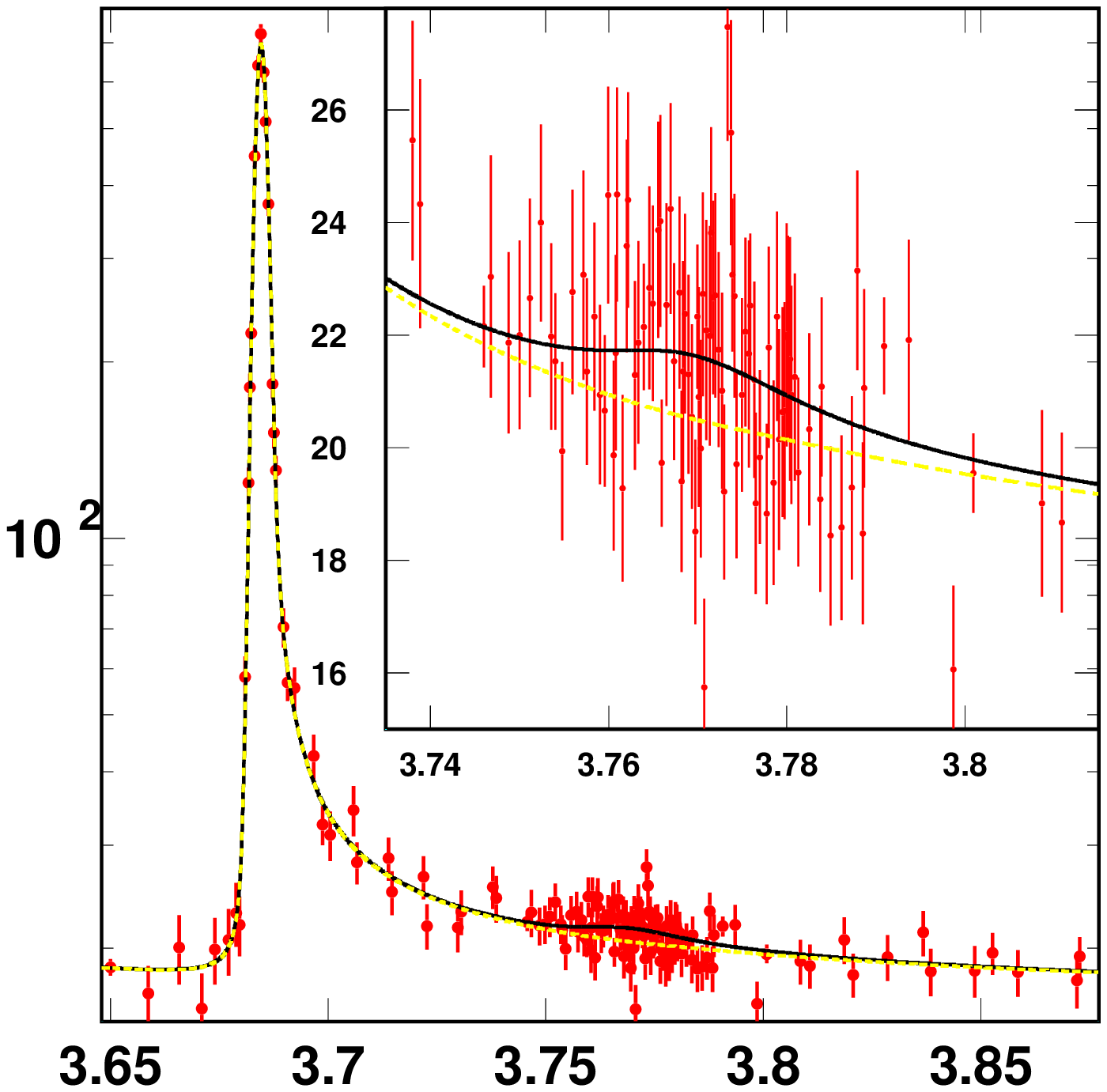}
\put(-140,-6.0){\bf\large $E_{\rm cm}$ [GeV]}
\put(-255,90){\rotatebox{90}
        {\bf\large $\sigma^{\rm obs}_{{\rm had-non}D\bar D}$~~~[nb]}}
\caption{The inclusive non-$D\bar D$ hadronic cross sections versus
the nominal c.m. energies (see text)~\cite{besii_psi3770_non_dd_3}.
}
\label{non_dd_xscts_vs_ecm}
\end{figure}

Table~\ref{tbl:bf_psi3770} summarizes the results for these measurements.
Among these measurements, the BES used the information about the direct measurements 
of the cross sections for $e^+e^- \rightarrow$ hadron$|_{{\rm non}-D\bar D}$ in the energy region
between 3.650 and 3.872 GeV to directly measure the branching fraction for $\psi(3770)\rightarrow$
non-$D\bar D$ for the first time, 
where {hadron$|_{{\rm non}-D\bar D}$} is the hadronic events 
which are not coming from the $D\bar D$ decays.
Figure~\ref{non_dd_xscts_vs_ecm} shows the measured non-$D\bar D$ cross sections VS the center-of-mass
of energy together with the best fit to the cross sections, 
where the enhancement of the non-$D\bar D$ cross sections 
around 3.770 GeV reflects the non-$D\bar D$
decays from the $\psi(3770)$ resonance~\cite{besii_psi3770_non_dd_3}.
This enhancement combining with the cross sections 
for $e^+e^-\rightarrow$hadron$|_{{\rm non}-D \bar D}$
measured at 3.773 and 3.650 GeV by the BES~\cite{besii_psi3770_non_dd_4} 
give a $4.8\sigma$ signal significance
for observing the signal of $\psi(3770)\rightarrow$ non-$D\bar D$ decays 
in the inclusive decay mode,
which gives the probability that the observed non-$D\bar D$ signal 
is due to the background fluctuation is about $2\times 10^{-6}$. 
The measured non-$D\bar D$ branching fraction of the $\psi(3770)$ decays obtained by analyzing
the cross sections for $e^+e^- \rightarrow$ hadron$|_{\rm non-}D\bar D$
exclude the possibility of that the measured non-$D\bar D$ decay branching fraction 
is due partially to the possible interference effects among the processes of
$\psi(3770) \rightarrow D\bar D$, continuum $e^+e^- \rightarrow D\bar D$ and
the $\psi(3686) \rightarrow D\bar D$ above the $D \bar D$  threshold.
Weighting these measured branching fractions for $\psi(3770) \rightarrow D^0\bar D^0$ and
$\psi(3770) \rightarrow D^+ D^-$ as listed in table~\ref{tbl:bf_psi3770}, 
the PDG gives the averaged branching fractions for these decays to be
$B[\psi(3770) \rightarrow D^0 \bar D^0]=(48.7 \pm 3.2)\%$,
$B[\psi(3770) \rightarrow D^+ D^-]=(36.1 \pm 2.8)\%$
and
$B[\psi(3770) \rightarrow D\bar D]=(85.3 \pm 3.2)\%$~\cite{pdg08},
leaving $(14.7\pm 3.2)\%$ of $\psi(3770)$ decay into non-$D\bar D$ final states.

Actually, with the $\psi(3770)$ reasonance parameters given by the PDG 
and the average of
the observed cross sections for $D\bar D$ production measured at 3.773 GeV, we can
determine the branching fraction for $\psi(3770)\rightarrow D\bar D$ and
for $\psi(3770)\rightarrow$ non-$D\bar D$ as well. 
This will be discussed in the next subsection.

\begin{table}
\caption{Measurements of branching fractions for
$\psi(3770)\rightarrow D \bar D$ and
for $\psi(3770)\rightarrow$ non-$D \bar D$ decays from the BES.}
\label{tbl:bf_psi3770}
\begin{tabular}{lccr} \hline \hline
$\psi(3770) \rightarrow$  &  $B (\%)$  &  $\psi(3770) \rightarrow$  & $B (\%)$  \\ \hline
$D^0\bar D^0$~\cite{besii_psi3770_non_dd_1}
              & $49.9 \pm 1.3 \pm 3.8$ &  $D \bar D$
              & $85.5 \pm 1.7 \pm 5.8$  \\
$D^+D^-$
              & $35.7 \pm 1.1 \pm 3.4$ &  non-$D \bar D$
              & $14.5 \pm 1.7 \pm 5.8$  \\ \hline
$D^0\bar D^0$~\cite{besii_psi3770_non_dd_2}
            & $46.7 \pm 4.7 \pm 2.3$ &  $D \bar D$  & $83.6 \pm 7.3 \pm 4.2$  \\
$D^+D^-$
              & $36.9 \pm 3.7 \pm 2.8$ &  non-$D \bar D$  & $16.4 \pm 7.3 \pm 4.2$  \\ \hline
$D\bar D$~\cite{besii_psi3770_non_dd_3}
              & $86.6 \pm 5.0 \pm 3.6$ &  non-$D \bar D$  & $13.4 \pm 5.0 \pm 3.6$  \\ \hline
$D\bar D$~\cite{besii_psi3770_non_dd_4}
              & $84.9 \pm 5.6 \pm 1.8$ &  non-$D \bar D$  & $15.1 \pm 5.6 \pm 1.8$  \\
\hline \hline
\end{tabular}
\end{table}
\begin{figure}
\includegraphics[width=9.5cm,height=5.5cm]
{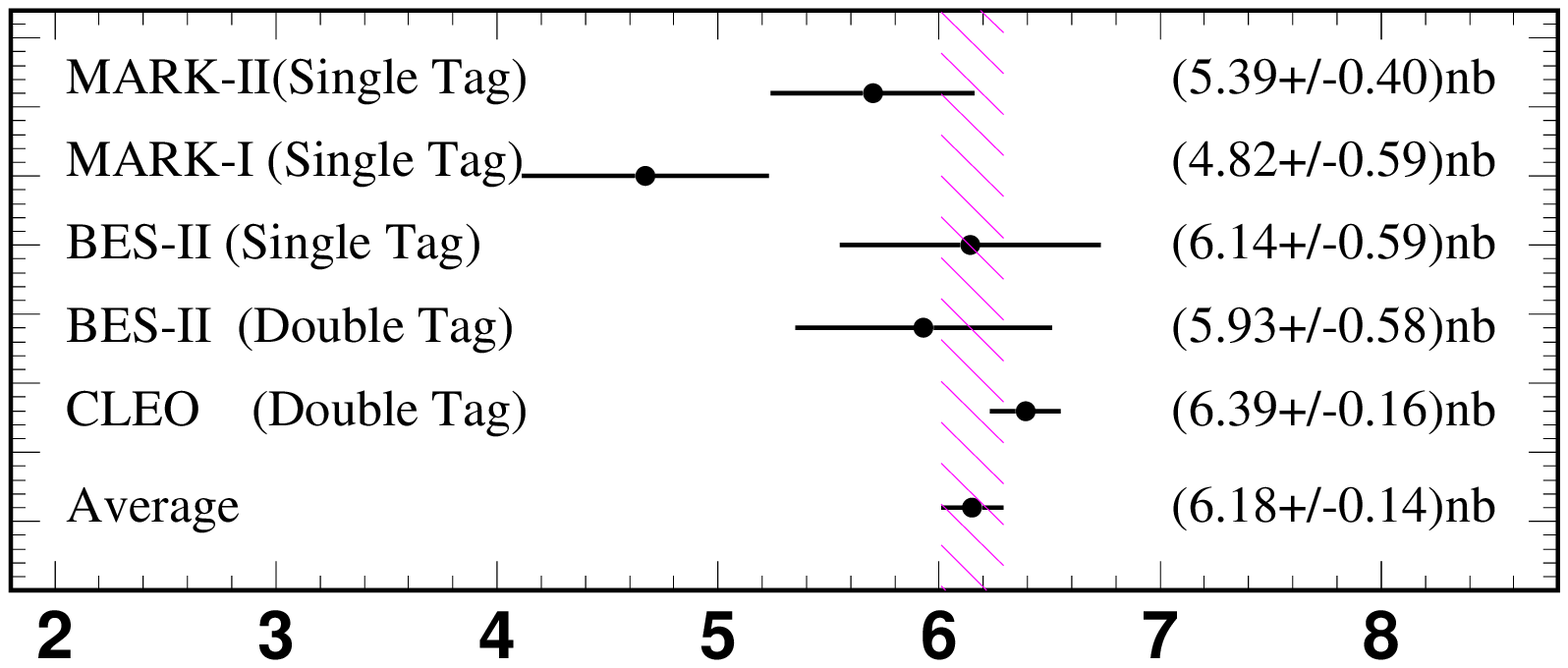}
\put(-160,5.0){\bf\large $\sigma^{\rm obs}_{D\bar D}$ [nb]}
\put(-165,42.0){\bf $\sigma^{\rm WA~obs}_{D\bar D}$ }
\caption{The observed cross section for $D \bar D$ production 
measured by different experiments at 3.771, 3.773 and 3.774 GeV (see text).}
\label{dd_xscts_3773mev}
\end{figure}

\subsection{$B[\psi(3770)\rightarrow {\rm non-}D\bar D]$ determined with the world averages 
of cross sections for $D\bar D$ and $\psi(3770)$ production}

Instead of directly measuring the non-$D\bar D$ branching fraction in the decays of $\psi(3770)$
in experiment, we can alternatively determine this non-$D\bar D$ branching fraction with the
$\psi(3770)$ resonance parameters given by the PDG and the observed cross section for $D\bar D$
production at 3.773 GeV. The observed cross sections for $D\bar D$ production measured by the BES and
CLEO Collaborations are consistent within error quite well (see figre~\ref{dd_xscts_3773mev}). 
The remained question
is what is the experimentally observed cross section for $\psi(3770)$ production at 3.773 GeV. 
The experimentlly observed cross section for $\psi(3770)$ production at 3.773 GeV 
can exactly be obtained in quantity based on the PDG
$\psi(3770)$ resonance parameters and ISR (Initial State Radiation) factor.
The experimentally observed cross section for $\psi(3770)$ production at 3.773 GeV is related to
production cross section by radiative corrections~\cite{kuraev, berends}.
These radiative corrections account for both the virtual photon effects and the real radiation which
reduce the actual center-of-mass energy down to lower energies. These effects effectively reduce the
observed cross section for $\psi(3770)$ production at 3.773 GeV. The amount of the reduction of
the observed cross section can be obtained based on the radiative corrections. 
The accuarcy in calculation of this amount is better than a few $0.1\%$.

If we believe that the PDG $\psi(3770)$ resonance parameters is right, 
we can make sure whether the BES measured non-$D\bar D$ branching fraction in the decays of $\psi(3770)$
is reliable.

Table~\ref{tbl:psi3770_parameters_pdg} lists the world averaged values of the $\psi(3770)$ resonance
parameters given by the  
PDG in 2006 and 2008. At the peak of the $\psi(3770)$,
the cross section for $\psi(3770)$ production is given by
\begin{equation}
\sigma^{\rm PRD}_{\psi(3770)}=\frac{12\pi \Gamma^{ee}}{M^2 \Gamma^{\rm tot}},
\end{equation}
where $M$ is the mass, $\Gamma^{\rm tot}$ is the total width and
$\Gamma^{ee}$ is the leptonic width of the $\psi(3770)$ resonance.
Inserting the PDG08 $\psi(3770)$ resonance parameters 
listed in table~\ref{tbl:psi3770_parameters_pdg}
to Eq.(1) yields the cross section for $\psi(3770)$ production
to be 
$$\sigma^{\rm PRD}_{\psi(3770)}=10.01\pm 0.68~~~{\rm nb}.$$
\noindent

At the $\psi(3770)$ peak, the initial state radiation factor is
$(1 + \delta)=0.770 \pm 0.014$~\cite{besii_psi3770_non_dd_2}. 
This ISR factor is used in calculation of the experimentally observed cross section 
for $\psi(3770)$ production at 3.773 GeV by the BES Collaboration~\cite{besii_psi3770_non_dd_2} 
with input of the PDG08 $\psi(3770)$ resonance parameters,
and in calculation of the $\psi(3770)$ leptonic width which corresponds to the
cross section for $\psi(3770)$ production~\footnote{The CLEO called
"the cross section for $\psi(3770)$ production" as "Born-level 
coss section~\cite{PRL96_2006_092002_CLEO}"} at 3.773 GeV
by the CLEO Collaboration~\cite{PRL96_2006_092002_CLEO} 
with input of the CLEO observed cross section 
$\sigma^{\rm obs}_{\psi(3770)}=6.38\pm 0.08^{+0.41}_{-0.30}$ nb~\cite{PRL96_2006_092002_CLEO}
for $\psi(3770)$ production. 
The CLEO gave the leptonic width of $\psi(3770)$ is 
$\Gamma^{ee}_{\psi(3770)}=204\pm 3^{+41}_{-27}$ eV~\cite{PRL96_2006_092002_CLEO},
which is 61 eV smaller than the $\Gamma^{ee}_{\psi(3770)}=265\pm 18$ eV~\cite{pdg08} 
given by the PDG08.
The CLEO $\Gamma^{ee}_{\psi(3770)}=204\pm 3^{+41}_{-27}$ eV
results in that the experimentally observed cross section for
$\psi(3770)$ productin measured by the CLEO at 3.773 GeV is
$1.54$ nb smaller than the experimentally observed cross section for
$\psi(3770)$ production at 3.773, which is extracted
with the $\psi(3770)$ parameters (see subsection D).

The MARK-I and the MARK-II Collaborations measured the $\sigma_{D}^{\rm obs}\cdot B$
near 3.773 GeV, where
$\sigma_{D}^{\rm obs}$ is the observed cross section for single $D^0$ or $D^+$ production, 
and $B$ is the branching fraction for $D^0$ or $D^+$ decay to the final state in question.
Table \ref{tab:dcrs-mark} lists the $\sigma_{D}^{\rm obs}\cdot B$ measured 
by the two Collaborations~\cite{mark1-dcrs,mark2-dcrs}. 
Using the PDG08 branching fractions for these decay modes~\cite{pdg08}, we can
obtain the observed cross sections for $D\bar D$ production (see figure~\ref{dd_xscts_3773mev}).

\begin{table}
\caption{$\sigma_{D}^{\rm obs}\cdot B$ measured by the MARK-I and the MARK-II collaborations.}
\label{tab:dcrs-mark}
\begin{tabular}{lcc} \hline \hline
Mode & $\sigma_{D}^{\rm obs}\cdot B$ & $\sigma_{D}^{\rm obs}\cdot B$\\
     & MARK-II [3.771 GeV]~~    & ~~MARK-I [3.774 GeV]            \\  \hline
$K^-\pi^+$          &$0.24\pm0.02$  &$0.25\pm0.05$ \\
$K^-\pi^+\pi^+\pi^-$&$0.68\pm0.11$  &$0.36\pm0.10$ \\
$K^-\pi^+\pi^+$     &$0.38\pm0.05$  &$0.36\pm0.06$ \\
\hline \hline
\end{tabular}
\end{table}

At the 3.773 GeV, the BES and CLEO
Collaborations previously measured the observed cross sections 
for $D\bar D$ production. Figure~\ref{dd_xscts_3773mev}
shows these measured cross sections and the observed
cross sections determined with $\sigma_{D}^{\rm obs}\cdot B$ measured by the MARK-I and the MARK-II
Collaborations. Weighting these observed cross sections for $D\bar D$ 
production measured at 3.773 GeV yields the world averaged value of the
observed cross sections for $D\bar D$ production to be
$\sigma^{\rm obs}_{D\bar D}|_{\sqrt{s}=3.773~{\rm GeV}} = (6.18 \pm 0.14)$ nb.
Assuming that there is no other new structure and effects except the $\psi(3770)$ resonance
in the energy range from 3.70 to 3.87 GeV, the branching fraction 
for $\psi(3770)\rightarrow D\bar D$ can be determined by~\cite{besii_psi3770_non_dd_2}

\begin{equation}
B[\psi(3770)\rightarrow D\bar D]=\frac{ \sigma^{\rm obs}_{D \bar D} }
                { (1 + \delta) \sigma^{\rm PRD}_{\psi(3770)}}.
\end{equation}
Here, we would like to stress the fact that both the $\sigma^{\rm obs}_{D\bar D}$
and $\sigma^{\rm PRD}_{\psi(3770)}$ are directly from experimental measurements in which
the effects from every possible dynamic or theoretical models have all been included in
both of them. So the ratio of the two cross sections can directly be used to determine the gap
between the cross section for $\psi(3770)$ production and the cross section for $D\bar D$ production,
at least to determine the $D\bar D$ or non-$D\bar D$ branching fractions of the $\psi(3770)$ decays
in any case.
 
Inserting these observed cross section, production cross section
and the ISR factor to Eq.(2) yields
the branching fraction for $\psi(3770)\rightarrow D\bar D$ to be
$$B[\psi(3770)\rightarrow D\bar D]=(80.2\pm 1.8 \pm 5.6)\%$$
\noindent 
and non-$D\bar D$ branching fraction of
$$B[\psi(3770)\rightarrow {\rm non}-D\bar D]=(19.8\pm 1.8 \pm 5.6)\%,$$
\noindent
where the first error is due to the uncertainty of the world average 
of the observed cross sections
for $D\bar D$ production at 3.773 GeV, the second arising from the uncertainties of
the ISR factor and the 
cross section for $\psi(3770)$ production at 3.773 GeV, which is calculated with the input
of the PDG08 $\psi(3770)$ resonance parameters.
These branching fractions are consistent within error 
with the ones (see table~\ref{tbl:bf_psi3770}) measured by the BES Collaboration.
Since we could not consider the correlations among the PDG $\psi(3770)$ parameters in
calculation of the cross section for $\psi(3770)$ production at 3.773 GeV,
the second error of $\pm 5.6\%$ in the determined branching fraction of $\psi(3770)$
is larger than the error of $\pm 3.2\%$ in the BES measured branching fractions.
In fact, the determined branching fraction for $\psi(3770)$ non-$D\bar D$ decays includes
all contributions from the DELCO, MARK-II, CLEO-c, and BES-II experiments, 
which contributed their measured cross sections for $\psi(3770)$ production to
the PDG $\psi(3770)$ resonance parameters~\cite{pdg08},
as well as the contribution from the MARK-I experiment which
contributed their oberseved cross section for $D\bar D$ production measured at 3.773 GeV
to the average (see Fig.~\ref{dd_xscts_3773mev}).
The large gap reflected by the ratio can not be remedied by any theoretical assumption
but only the non-$D \bar D$ decays of $\psi(3770)$.

Alternatively, if using the PDG06 $\psi(3770)$ resonance parameters to calculate the
observed cross section for $\psi(3770)$ production at its peak, 
we obtain the non-$D\bar D$ decay branching fraction
to be $B[\psi(3770)\rightarrow {\rm non}-D\bar D]=(26.5\pm 1.7 \pm 11.7)\%$. 
The PDG06 $\psi(3770)$ resonance
parameters did not include the BES-II measurements of the parameters.

\begin{table}
\caption{The world averaged $\psi(3770)$ resonance parameters given by the Particle Data Group,
where $M$ is the mass,
$\Gamma^{\rm tot}$ is the total width,
and $\Gamma^{ee}$ is the partial leptonic width of the resonance~\cite{pdg08}.}
\label{tbl:psi3770_parameters_pdg}
\begin{tabular}{lccr} \hline \hline
M (MeV)~~~~~~~  &  ~~~~$\Gamma^{\rm tot}$ (MeV)~~~~  & ~~~~$\Gamma^{ee}$ (eV)~~~~ & ~~~~Note
                                                     \\  \hline
$3772.92\pm 0.35$  &        $27.3 \pm 1.0$        &  $265 \pm 18$           &    PDG08  \\
$3771.1\pm 2.4$    &        $23.0 \pm 2.7$        &  $242^{+27}_{-24}$      &    PDG06  \\
\hline \hline
\end{tabular}
\end{table}

\subsection{Other analysis of the cross sections for $D\bar D$ production in open-charm energy region}

Recently, Li (Li, Qin and Yang) published a paper entitled "Study of the branching ratio of 
$\psi(3770) \rightarrow D\bar{D}$ in $e^+e^- \rightarrow D\bar{D}$ 
scattering"~\cite{PRD81_2010_p011501_LiHB}, 
and declared that their result is different from that of BES Collaboration. 
However their analysis technique and their result suffered from some serious problems which
lead to that their analysis can not give correct result on $\psi$(3770) and 
other resonance decays. We here list the problems as follows:

(1) In their analysis, they assumed that the $D\bar{D}$ production cross section 
can be described with a square of the sum of coherent amplitudes of resonances 
above $D\bar{D}$ threshold, each of which is described by a Beit-Wigner amplitude and has 
own mass, total width and leptonic width. These mass, total width and leptonic 
width should be obtained from fitting their own formula to the experimental 
data directly. However, the resonance parameters of $\psi(3770)$, $\psi(4040)$ and 
$\psi(4160)$ in their fits were all taken directly from the PDG values
of the parameters. These parameters in the PDG were all obtained in the simply 
fundamental physical assumption. In the experiments to measure these 
parameters, people did not consider the effects of $\psi$(3686) and the 
interferences among $\psi(3686)$, $\psi(3770)$, $\psi(4040)$ and $\psi(4160)$
on the measured values of the resonance parameters. At  present, all of 
the resonance parameters for $\psi(3770)$, $\psi(4040)$, $\psi(4160)$ were obtained 
without considering these effects. If $\psi$(3686) really affects $\psi(3770)$ 
production and decays as large as the one reported in their paper, these resonance 
parameters given in the PDG could not be used and fixed
in their analysis. Instead they have to leave 
these resonance parameters free in their fits to the data to obtain
the decay branching fractions and resonance parameters such as the mass $M^{\rm Li}$, 
the total width $\Gamma^{\rm Li}$,
and the leptonic width $\Gamma^{\rm Li}_{ee}$ for these resonances. 
If they fix these resonance parameters at the values given in the PDG in their fits, 
they cannot give correct results on measurements
of these branching fractions for $\psi(3770)$, $\psi(4040)$ and $\psi(4160)$ decays
to $D\bar D$ or to non-$D\bar D$ since the $\Gamma^{\rm Li}_{ee}[\psi(3770)]$ and other 
parameters$^{\rm Li}$ are not indentical to the $\Gamma_{ee}[\psi(3770)]$ 
and the other parameter values given in the PDG. 

(2) They not only have serious problems with the inputs of their fixed resonance 
parameters, but also have a problem with handling their results from their fits.
In their analysis, they totally obtained eight different solutions from their fits.
However, they dropped six of the eight solutions because these six solutions 
give the $\psi(3770)$ non-$D\bar{D}$ decay branching fractions to be larger than 30\%. 
They only retained two solutions which give 
$\psi(3770)$ non-$D\bar{D}$ decay branching fractions are $(2.8\pm 8.9)\%$ and 
$(-1.1\pm 9.0)\%$.  Based on the two retained solutions, they declared that 
their fitted result is different from that of BES Collaboration. 

(3)Moreover, according to their two retained solutions, they obtained the 
branching fractions for $\psi(4040) \rightarrow D\bar{D}$ decays 
are $(25.3 \pm 4.5)\%$ and $(34.7\pm 4.8)\%$, 
which largely deviate from the values observed in $e^+e^-$
experiments. For example, from the analysis of the
data taken at 4.03 GeV with the BES-I detector, from the cross sections for
$D\bar{D}$, $D\bar{D^*}$ and $D^*\bar{D^*}$ production measured 
by the CLEO-c~\cite{Blusk_CLEO_c_CP870} 
and from these 
cross sections predicted by Eitchten's Couple-Channel Model, one knows that the branching fraction 
for $\psi(4040) \rightarrow D\bar{D}$ is only less than 3\%. 
These also indicate that their analysis did not give correct results on $\psi$(3770), $\psi(4040)$
and $\psi(4160)$ decays.

Ignoring 
above serious problems which
lead to that their analysis can not give correct result on $\psi(3770)$
non-$D\bar D$ decays,
one can clearly find that 
their $B[\psi(3770)\rightarrow {\rm non-}D\bar D]=(2.8\pm 8.9)\%$ 
is consistent within error
with $B[\psi(3770)\rightarrow {\rm non-}D\bar D]=(14.7\pm 3.2)\%$ measured by the BES.
So, from Li's branching fraction of $\psi(3770)$ decays,
no one can claim that Li's result on the branching fraction
for $\psi(3770)$ non-$D\bar D$ decays is different with that of the BES.
In fact,
Li's analysis can not give any significant conclusion
on whether the non-$D\bar D$ branching fraction of
$\psi(3770)$ decays is in the range from $-7\%$ to $12\%$
or out the range.
It never concludes that $\psi(3770)$ does not decay into
non-$D\bar D$ final states with a branching fraction of about $14\%$
like that measured by the BES.
If one want to give a definite conclusion on the non-$D\bar D$ decays of
$\psi(3770)$ with Li's branching fraction of $\psi(3770)$ decays,
the $B[\psi(3770)\rightarrow {\rm non-}D\bar D]=(2.8\pm 8.9)\%$
may only indicate that the upper limit of the non-$D\bar D$
branching fraction of $\psi(3770)$ decays is less than about $20\%$ at $90\%$ C.L..

\subsection{Cross sections for $\psi(3770)\rightarrow$ non-$D\bar D$ measured at
the CLEO-c and the BES-II}

Instead of measuring the non-$D\bar D$ branching fraction of $\psi(3770)$ decay, 
the CLEO Collaboration measured the cross section for $\psi(3770)\rightarrow$ non-$D\bar D$ based on
analyzing their data taken at 3.773 and 3.671 GeV. The BES Collaboration also measured 
the cross section for $\psi(3770)\rightarrow$ non-$D\bar D$ decays by analyzing the data
taken at 3.773, 3.650 GeV, and the data taken in the energy range from 3.650 to 3.872 GeV. 
The second column of table~\ref{tbl:non_dd_xscts_psi3770}
shows the non-$D\bar D$ cross sections,
where the first three rows of the table
summarize the results of these measurements, while the 
fourth and fifth rows list the
non-$D \bar D$ cross sections obtained 
by a simple calculation based on the observed cross sections for
$\psi(3770)$ production~\cite{besii_psi3770_non_dd_1}, 
for $D\bar D$ production~\cite{xsct_dd_at_3770mev} 
and the non-$D\bar D$ branching fraction~\cite{besii_psi3770_non_dd_2} 
for $\psi(3770)$ decays from the BES. 
The third column of the table
shows the observed cross sections for $\psi(3770)$ production measured at 3.773 GeV.

Actually, $(1+\delta)\sigma^{\rm PRD}_{\psi(3770)}$ gives the world average
of the experimentally observed cross sections for $\psi(3770)$ production at 3.773 GeV,
while $\sigma^{\rm obs}_{D\bar D}/(1+\delta)$
gives the cross section for $D\bar D$ production at 3.773 GeV.
One can directly either compare the two
experimentally observed cross section or the two production cross sections at 3.773 GeV
to measure the non-$D\bar D$ branching fraction in the decays of $\psi(3770)$. 
With the ISR factor $(1+\delta)$ and 
$\sigma^{\rm PRD}_{\psi(3770)}$, we obtain the world average
of the experimentally observed cross sections for
$\psi(3770)$ production at 3.773 GeV to be
$$\sigma^{\rm WA~obs}_{\psi(3770)} = (7.71\pm 0.52 \pm 0.14)~~~ {\rm nb},$$
\noindent
which is consistent within error with these
(see table~\ref{tbl:non_dd_xscts_psi3770}) measured by the BES and the one
measured by the MARK-II (see table~\ref{tbl:non_dd_xscts_psi3770}), 
but more than $2\sigma$ larger
than $6.38\pm 0.08^{+0.41}_{-0.30}$ measured by the CLEO.
With $\sigma^{\rm WA~obs}_{\psi(3770)} = (7.71\pm 0.52 \pm 0.14)$ nb
and $\sigma^{\rm WA~obs}_{D\bar D} = (6.18\pm 0.14)$ nb of
the world average of the observed cross sections for $D\bar D$ 
production (see Fig.~\ref{dd_xscts_3773mev}), 
we obtain
the world average of cross sections
for $\psi(3770)\rightarrow$non-$D\bar D$ decays to be
$$\sigma^{\rm WA~obs}_{{\rm non}-D\bar D}=(1.53\pm 0.52\pm 0.20)~~~{\rm nb},$$
\noindent
which is consistent within error with these (see table~\ref{tbl:non_dd_xscts_psi3770})
measured by the BES, but 1.54 nb larger than
the one (see table~\ref{tbl:non_dd_xscts_psi3770}) measured by the CLEO.

\begin{table}
\caption{Measurements of non-$D\bar D$ cross sections for
$\psi(3770)$ decays and the experimentally observed cross section 
for $\psi(3770)$ production at 3.773 GeV,
where 'WA and PDG08' indicate that the quantities listed in the row are obtained with the input of
the world average of the observed cross sections for $D\bar D$ production at 3.773 GeV
and the inputs of the PDG08 $\psi(3770)$ resonance parameters as well as the
ISR factor (see text).}
\label{tbl:non_dd_xscts_psi3770}
\begin{tabular}{lcr} \hline \hline
Experiment  &  $\sigma^{\rm obs}_{{\rm non}-D\bar D}$ [nb]
            &  $\sigma^{\rm obs}_{\psi(3770)}$ [nb]  \\ \hline
CLEO-c~\cite{PRL96_2006_092002_CLEO} &  $-0.01\pm 0.08^{+0.41}_{-0.30}$ & $6.38\pm 0.08^{+0.41}_{-0.30}$ \\
BES-II~\cite{besii_psi3770_non_dd_1} &  $1.14\pm 0.08\pm 0.59$ & $7.18\pm 0.20 \pm 0.63$ \\
BES-II~\cite{besii_psi3770_non_dd_2} &  $1.04\pm 0.23\pm 0.13$ & $6.94\pm 0.48 \pm 0.28$         \\
BES-II~\cite{besii_psi3770_non_dd_3} &  $0.95\pm 0.35\pm 0.29$ & $7.07\pm 0.36 \pm 0.45$         \\
BES-II~\cite{besii_psi3770_non_dd_4} &  $1.08\pm 0.40\pm 0.15$ &      ---        \\
MARK-II~\cite{markii_psi3770_prd21_p2716} & -- & $9.1\pm 1.4$  \\
WA and PDG08                              & $1.53\pm 0.52 \pm 0.20$ & $7.71\pm 0.52\pm 0.14$ \\
\hline \hline
\end{tabular}
\end{table}

\section{Light Hadron Decays}

Summing over all of the measured non-$D\bar D$ decay branching fractions
for the hadronic and electromagnetic transitions of $\psi(3770)$
gives a totally measured non-$D\bar D$ decay branching fractions of $\psi(3770)$
to be about $2\%$. Comparing the measured total branching fractions of these exclusive
decays with the total inclusive non-$D\bar D$ branching fraction of $\psi(3770)$ decays
measured by the BES, we find that
about another $12\%$ of $\psi(3770)$ non-$D\bar D$ decays have not been found yet.
To find the light hadron decays of $\psi(3770)$, both the CLEO and the BES Collaborations
extensively studied the possible light hadron decay modes of $\psi(3770)$.

\subsection{$\psi(3770)\rightarrow \phi\eta$ decay}

Both the BES and CLEO Collaborations searched for more
exclusive non-$D\bar D$ decay processes of $\psi(3770)$ from their data taken
at 3.773 GeV and at 3.650 or 3.671 GeV. 
Up to now,
the BES~\cite{BES_LH_papers} and the CLEO~\cite{CLEO_LH_papers,PRD73_2006_p012002_cleo} 
have searched for more than 60 light hadron 
decay modes
for $\psi(3770) \rightarrow {\rm LH}$ (LH is light hadron), 
but they did not claim significant signal events for these decays, except
the decay $\psi(3770)\rightarrow \phi\eta$.

In 2006, the CLEO claimed that they had found the light hadron decay for
$\psi(3770)\rightarrow \phi\eta$ and measured the branching fraction~\cite{PRD73_2006_p012002_cleo}
$$B[\psi(3770)\rightarrow \phi\eta]=(3.1\pm 0.6\pm 0.3)\times 10^{-4}.$$
\noindent
This branching fraction is obtained with the net cross section for
$e^+e^- \rightarrow \phi\eta$ measured at 3.773 GeV,
which is the difference between the cross section measured at 3.773 GeV
and the one measured at 3.671 GeV. In the determination of this branching fraction,
the CLEO ignored the possible interference among the amplitudes for this final state from
the $\psi(3686)$ and the $\psi(3770)$ decays as well as from the continuum production in
$e^+e^-$ annihilation.
  
\subsection{Discussion on $\psi(3770)\rightarrow {\rm LH}$ decays}

Although the CLEO did not claim observations for other light hadron decay modes
of $\psi(3770)\rightarrow {\rm LH}$, from their measurements of the cross sections,
if assuming that there is only one $\psi(3770)$ resonance in the range from 3.70 to 3.87 GeV,
one still can find that there are some strong evidences for existing the light hadron decays
of the $\psi(3770)$ resonance.

Table~\ref{tbl:LH_xscts_3772mev_3671mev_CLEO} lists some cross sections
for $e^+e^-\rightarrow {\rm LH}$ measured at 3.773 and 3.671 GeV by 
the CLEO~\cite{PRD73_2006_p012002_cleo}.
From this table, 
we can find that
the values of the cross section for 
$e^+e^- \rightarrow \pi^+\pi^-\pi^0,\rho\pi,\omega\eta$
measured at 3.773 GeV is significantly
lower than the ones measured at 3.671 GeV,
while the cross section for $e^+e^- \rightarrow \phi\eta$ measured at 3.773 GeV
is higher than the one measured at 3.671 GeV.
In addition, the cross sections for $e^+e^- \rightarrow K^{*0}\bar {K^0}$ are
larger than the ones for $e^+e^- \rightarrow K^{*+}{K^-}$ by a factor of 23 at
the two energies. 
These indicate that there must be some dynamics effects which destroy the
$\pi^+\pi^-\pi^0$, $\rho\pi$, and $\omega\eta$ production from the $\psi(3770)$ decays,
but these effects enhance the $\phi\eta$ production from the $\psi(3770)$ decays.
The physical effects destroying or enhancing these channel production from the $\psi(3770)$
decays also
affect other channel production from the $\psi(3770)$ decays, resulting in no significant signal
events for $\psi(3770)\rightarrow {\rm LH}$ observed by directly looking at the cross sections 
for these channel production at the two energies of 3.773 GeV and 3.671 GeV.
In addition, asymmetry production for $e^+e^-\rightarrow K^{*0}\bar {K^0}$ and
for $e^+e^-\rightarrow K^{*+}{K^-}$ at the two energies also indicates 
that there must be some dynamics effects affecting the strange meson production.
The simple explanation for these destroyed or enhanced cross sections measured at 3.773 GeV
is due to the interference among the amplitudes for $\psi(3686)$ and $\psi(3770)$ 
decays to these final states as well as the amplitudes for these final states 
produced in $e^+e^-$ annihilation directly. 

There is also another possibility to explain these measurements of the cross sections. 
The BES Collaboration observed an anomalous line shape of the cross sections for
$e^+e^- \rightarrow {\rm hadron}$ in the energy region 
between 3.70 and 3.87 GeV~\cite{line_shape_hadrons}.
This anomalous line shape of the cross sections was interpreted as an Di-resonances by
Dubynskiy and Voloshin~\cite{PRD78_2008_p116014_Voloshin_2008}, 
which is the same as the possible new structure 
claimed in the BES published
paper~\cite{line_shape_hadrons}. 
If there is really existing the new structure R(3765) near 3.765 GeV, 
and assuming that this structure
can decay into these final states, the increased and the decreased cross sections for these
final states can be explained as the results of the interference among these amplitudes.

Due to the lower statistics, the BES Collaboration did not find significant differences of
the observed cross sections for the exclusive light hadronic event production
at 3.773 and 3.650 GeV. The CLEO Collaboration find some significant differences for some channels
for $e^+e^- \rightarrow {\rm hadron}$. However, the CLEO only claimed they find the light
hadron decay mode for $e^+e^-\rightarrow \phi\eta$ since they measured the cross section for
$e^+e^-\rightarrow \phi\eta$ at 3.773 GeV being $2.4^{+2.0}_{-1.3}$ nb
larger than the one at 3.671 GeV. But, they ignored the decay modes for 
$e^+e^-\rightarrow \pi^+\pi^-\pi^0$, $e^+e^-\rightarrow \rho\pi$ and
$e^+e^-\rightarrow \omega\eta$, for which the cross sections measured
at 3.773 GeV are, respectively, smaller $-5.7^{+1.9}_{-1.7}$, 
$-3.6^{+1.7}_{-1.4}$ and $-1.9^{+1.8}_{-1.0}$ than the ones at 3.671 GeV.
These negative net cross sections measured at 3.773 GeV also strongly suggest that the
$\psi(3770)$ do decay into these final states or there are some dynamic effects destroying
these processes of $\psi(3770)$ decays.

In the simplest case of assuming that there is no new structure in the energy
range from 3.70 to 3.87 GeV and there is no new dynamics effects affecting the $\psi(3770)$
production and decays, one can extract out the decay branching fractions for these
final states from the $\psi(3770)$.
To find the light hadron decays of $\psi(3770)$ and measure these decay branching fractions, 
one may have to make a global analysis 
of the cross sections for different channel production measured at the two energy points
with considering the possible interference among the decay amplitudes for each process 
at least. 

In 2004, the BES Collaboration observed a large cross section for 
$e^+e^-\rightarrow K^{*0}\bar {K^0}+c.c.$ production, 
which is $\sigma(e^+e^- \rightarrow K^{*0}\bar {K^0}+c.c.)=(15.0\pm 4.6\pm 3.3)$ pb 
at 3.773 GeV, and found that the process for $e^+e^-\rightarrow K^{*\pm}(892)K^{\mp}$ 
is much suppressed. These observations are confirmed 
by the CLEO measurements for the same channels 
(see table~\ref{tbl:LH_xscts_3772mev_3671mev_CLEO}).
Taking into account the possible interference between the strong decay amplitude and the
continuum production amplitude at 3.773 GeV, the BES Collaboration 
measured the decay branching fraction
for $\psi(3770) \rightarrow K^{*0}\bar{K^0}+c.c.$ to be 
$$B[\psi(3770) \rightarrow K^{*0}\bar{K^0}+c.c.]=(4.3^{+5.4}_{-3.4}\pm 1.3)\times 10^{-4},$$
\noindent
set the upper limits on the strong decay branching fraction and
the partial width~\cite{RongG_ICHEP04} to be
$$B[\psi(3770) \rightarrow K^{*0}\bar {K^0}+c.c.)]<0.12\%~~~~{\rm at~}90\%~~{\rm C.L.} $$
\noindent
and
$$\Gamma[\psi(3770) \rightarrow K^{*0}\bar {K^0}+c.c.)]<29~~{\rm keV}~~~~~{\rm at~}90\%~~{\rm C.L.},$$
\noindent
respectively.

Recently, D. Zhang developed a model~\cite{D_Zhang_hep_ex_0808.0091} to incorporate the decays 
of $\psi(3770)\rightarrow {\rm VP}$ (Vector Pseudoscalar). With the observed cross sections 
measured at 3.773 and 3.671 GeV by the CLEO Collaboration, 
he found that the branching fraction for $\psi(3770)\rightarrow \rho\pi$ to be 
$B[\psi(3770)\rightarrow \rho\pi]=(0.183^{+0.061}_{-0.067})\%$, 
which is within the range from $6\times 10^{-6}$ to $2.4\times 10^{-3}$ measured 
by the BES~\cite{PRD72_2005_p072007}.
This branching fraction correspond to
the partial width of $\Gamma[\psi(3770)\rightarrow \rho\pi]=49.7^{+16.9}_{-18.3}$ 
keV~\cite{D_Zhang_hep_ex_0808.0091}.
In $J/\psi$ decays, the fraction of the partial width for $J/\psi \rightarrow \rho\pi$ to
the partial width for strong decays of $J/\psi \rightarrow {\rm hadron}$ is
$$f_{J/\psi}=\frac{\Gamma(J/\psi \rightarrow \rho\pi)}
                  {\Gamma(J/\psi \rightarrow {\rm hadron})_{3g}}=0.017.$$
If assuming that 
the light hadron decays of the $\psi(3770)$ go through 3 gloun annihilation and
the fraction $f_{\psi(3770)}$ of the partial width 
for $\psi(3770)\rightarrow \rho\pi$ to the partial width for $\psi(3770) \rightarrow {\rm LH}$ 
is roughly as the same as $f_{J/\psi}$ for $J/\psi$ strong decays,
the partial width for $\psi(3770)\rightarrow \rho\pi$ indicates that 
$(10.8^{+3.6}_{-3.9})\%$
of $\psi(3770)$
decays to light hadron final states. Considering about $2\%$ of $\psi(3770)$ hadronic and $\gamma$
transitions, the total non-$D\bar D$ decay branching fraction of $\psi(3770)$ would be as large
as $(12.8^{+3.6}_{-3.9})\%$,
which is as the same as 
$B[\psi(3770)\rightarrow {\rm non-}D\bar D]=(14.7\pm 3.2)\%$ 
measured by the BES Collaboration.

\begin{table}
\caption{Measurements of cross sections for $e^+e^- \rightarrow \pi^+\pi^-\pi^0$ 
and $e^+e^- \rightarrow {\rm VP}$ channels
at 3.773 and 3.671 GeV by the CLEO~\cite{PRD73_2006_p012002_cleo}.} 
\label{tbl:LH_xscts_3772mev_3671mev_CLEO}
\begin{tabular}{lcr} \hline \hline
Channel~~~~~~~  &  ~~~~$\sigma^{3.671 {\rm GeV}}$ [pb]~~~~  & ~~~~$\sigma^{3.773 {\rm GeV}}$ [pb]~~~~
                                                     \\  \hline
$\pi^+\pi^-\pi^0$  &        $13.1^{+1.9}_{-1.7} \pm 2.1$   & $7.4 \pm 0.4 \pm 2.1$ \\
$\rho\pi$          &        $ 8.0^{+1.7}_{-1.4} \pm 0.9$   & $4.4 \pm 0.3 \pm 0.5$ \\
~~~$\rho^0\pi^0$   &        $ 3.1^{+1.0}_{-0.8} \pm 0.4$   & $1.3 \pm 0.2 \pm 0.2$ \\
~~~$\rho^+\pi^-$   &        $ 4.8^{+1.5}_{-1.2} \pm 0.5$   & $3.2 \pm 0.3 \pm 0.2$ \\
$\omega\eta$       &        $ 2.3^{+1.8}_{-1.0} \pm 0.5$   & $0.4 \pm 0.2 \pm 0.1$ \\
$\phi\eta$         &        $ 2.1^{+1.9}_{-1.2} \pm 0.2$   & $4.5 \pm 0.5 \pm 0.5$ \\
$K^{*0}\bar {K^0}$ &        $ 23.5^{+4.6}_{-3.9}\pm 3.1$   & $23.5\pm 1.1 \pm 3.1$ \\
$K^{*+}{K^-}$      &        $  1.0^{+1.1}_{-0.7}\pm 0.5$   & $<0.6$ \\
\hline \hline
\end{tabular}
\end{table}

\subsection{Some theoretical predictions for the non-$D \bar D$ decays}

\subsubsection{Assuming $\psi(3770)$ being pure $c\bar c$ state}
With considering the $\psi(3770)$ as a pure $c\bar c$ state, 
some theoretical physicists calculated the non-$D\bar D$ branching fraction of $\psi(3770)$ decays.

In 2008, by introducing the color-octet mechanism calculated up to
next to leading order within the framework of NRQCD, He, Fan and Chao calculated
the light hadron branching fraction of $\psi(3770)$ decays~\cite{GT_Chao_NRQCD_cal}. 
They reported that the light hadron
branching fraction of $\psi(3770)$ decays is $(2.0^{+1.5}_{-0.8}\pm 1.0)\%$ and 
said that it could be as large as $(3.5\pm 1.8)\%$. Considering the measured branching fractions
for the hadronic and electromegnatic transitions of $\psi(3770)$, they claimed 
the maximum value of the non-$D\bar D$ branching fraction of $\psi(3770$ decay is $5\%$.
In their published paper, 
they pointed out that the new decay mechanism has to be considered if the
non-$D\bar D$ decay branching fraction is significantly larger than $5\%$~\cite{GT_Chao_NRQCD_cal}.

In May 2008, Liu, Zhang and Li calculated the non-$D\bar D$ branching fraction of $\psi(3770)$ decays
by taking final state interaction (FSI) into account in the decays~\cite{Liu_Zhang_Li_PLB675_2009_p441}. 
They found the contribution to the
non-$D\bar D$ decay branching fraction from the final state interaction can reach up to
$B^{\rm FSI}_{{\rm non}-D\bar D} = (0.2-1.1)\%$. Adding the total contribution of the NRQCD and FSI
yields the upper band of the light hadron branching fraction to be up to $4.6\%$. 
Combining the $4.6\%$ and the branching fractions for hadronic and electromegnatic transitions,
they gave the upper band of the non-$D\bar D$ branching fraction of $\psi(3770)$ decays to be
$6.4\%$~\cite{Liu_Zhang_Li_PLB675_2009_p441}. 

In May 2008, Zhang, Li and Zhao 
studied the intermediate hadron exchange process in the $\psi(3770)$
VP decays~\cite{PRL102_2009_172001_Zhao}. They found the decay branching fraction 
for $\psi(3770)\rightarrow {\rm VP}$ is in the range from $0.41\%$ to $0.64\%$, 
comparing about $0.3\%$ of the sum of branching fractions for 10 modes of 
$\psi(3770) \rightarrow {\rm VP}$ 
decays obtained by D. Zhang~\cite{D_Zhang_hep_ex_0808.0091}.
Zhang, Li and Zhao pointed out in Ref.~\cite{PRL102_2009_172001_Zhao} 
that the long-range interactions play a role in $\psi(3770)$ strong
decays, and said it could be a key towards a full understanding of the mysterious $\psi(3770)$
non-$D\bar D$ decay mechanism~\cite{PRL102_2009_172001_Zhao}.

\subsubsection{Other explanations for large non-$D\bar D$ decays of $\psi(3770)$}

In 2005, J. Rosner proposed a model of reannihilation of the $D\bar D$ pair to explain
the non-$D\bar D$ decays of $\psi(3770)$~\cite{J_Rosner_hep_ph_0411003}.

The large non-$D\bar D$ branching fraction in the decays of $\psi(3770)$ measured by the BES
can be explained with a suggestion of a sizeable four-quark component 
in addtion to the $c\bar c$ state.
M.B. Voloshin suggested that "the $\psi(3770)$ resonance may contain a sizeable [$\O(10\%)$
in terms of the probability weight factor] four-quark component with the up- and down-quarks and
antiquarks in addition to the $c\bar c$ pair, 
which component in itself has a substantial part with isospin $I=1$".
With his suggested
four-quark component of the wave function of the $\psi(3770)$, he expected
that the non-$D\bar D$ branching fraction
for $\psi(3770)$ decays is around 
$10\%$~\cite{BM_Voloshin_paper_prd71_2005_p114003},
and that the decay branching fraction for the hadronic transition of
$\psi(3770)\rightarrow J/\psi \eta$ is
$B[\psi(3770)\rightarrow J/\psi \eta] \sim 0.15\%$.
His prediction for the hadronic transition rate
was confirmed by the CLEO measurement, which gave
$B[\psi(3770)\rightarrow J/\psi \eta]=(0.087\pm 0.033\pm 0.022)\%$~\cite{PRL96_2006_082004_CLEO}.
In 2008, in his reviewing charmonium, M.B. Voloshin pointed out that, 
if the non-$D\bar D$ branching fraction of $\psi(3770)$
decays is significantly larger than the sum of the branching fractions 
for hadronic and electromagnetic transitions of $\psi(3770)$, 
it would imply an enhanced light hadron decays of $\psi(3770)$.
Such enhanced decays can be attributted to a presence of a certain four-quark component
in the wave function of $\psi(3770)$~\cite{M_B_Voloshin_Prog_Part_Nucl_Phys_61_2008_p455}.

\section{Summary}

\begin{figure}
\includegraphics[width=9.5cm,height=5.5cm]
{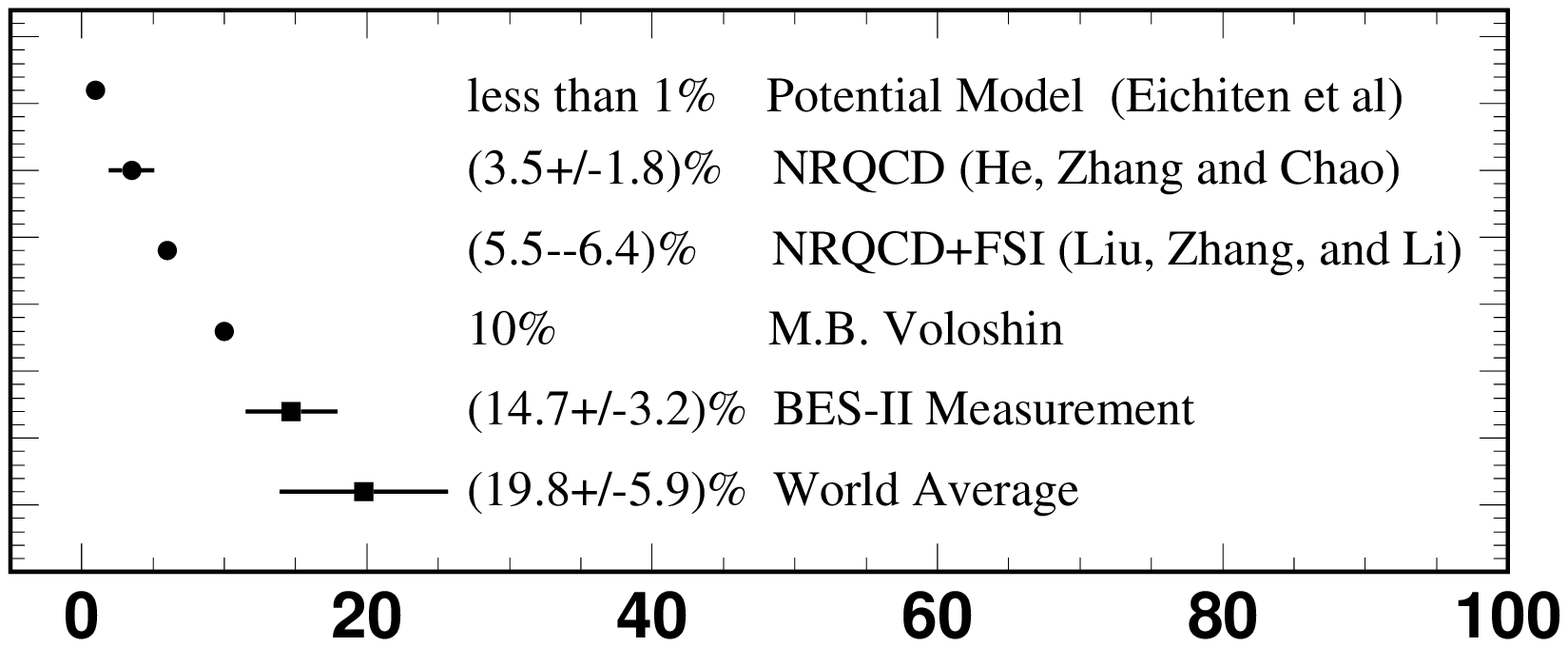}
\put(-180,5.0){\bf $B[\psi(3770)\rightarrow {\rm non-}D\bar D]$ [$\%$]}
\caption{The predicted and the measured non-$D\bar D$ branching fraction in the decays of $\psi(3770)$, 
where the World Average branching fraction is determined with the average of the observed cross sections
for $D\bar D$ production measured at 3.773 GeV by the MARK-I, MARK-II, BES-II and CLEO-c experiments, 
and the PDG08 $\psi(3770)$ resonance parameters (see text).}
\label{Br_non_DD_psi3770}
\end{figure}

In summary, we experimentally reviewed the progress on experimental studies of the $\psi(3770)$
non-$D\bar D$ decays. 
The determined $\psi(3770)$ non-$D\bar D$ decay branching fraction
$B[\psi(3770) \rightarrow {\rm non-}D\bar D]=(19.8 \pm 1.8 \pm 5.6)\%$,
which was obtained with the world averaged $\psi(3770)$ resonance parameters
and the world average of the observed $D\bar D$ cross section measured at 3.773 GeV,
is consistent within error with
the BES previously measured inclusive non-$D\bar D$ decay branching fraction, 
$B[\psi(3770) \rightarrow {\rm non-}D\bar D]=(14.7 \pm 3.2)\%$.
Figure~\ref{Br_non_DD_psi3770} shows a comparison 
of the previously measured branching fraction by the BES,
the determined branching fraction with the world average of the observed $D\bar D$
cross sections at 3.773 GeV and the PDG08 $\psi(3770)$ resonance parameters.
In order to directly compare the measured branching fractions for the non-$D\bar D$
decays of $\psi(3770)$ with the theoretical predictions for this decay branching fraction,
we also plot some theoretical predictions in the figure.
Both the BES previously measured and the determined world averaged 
inclusive non-$D\bar D$ branching fraction of $\psi(3770)$ decays
are significantly larger than the theoretical predictions 
of the potential model, the NRQCD calculation and the NRQCD+FSI calculations.
All of these predictions are based on the assumption of that $\psi(3770)$ is pure $c\bar c$ state.
However,
the BES measured inclusive non-$D\bar D$ decay branching fraction
$B[\psi(3770) \rightarrow {\rm non-}D\bar D]=(14.7 \pm 3.2)\%$
is more close to the $10\%$ expected by M.B. Voloshin based on his assumption of
four-quark component in addtion to $c\bar c$ state~\cite{BM_Voloshin_paper_prd71_2005_p114003}.
So, the large inclusive non-$D\bar D$ decay branching fraction 
in the decays of $\psi(3770)$ favour the assumption of that the $\psi(3770)$ resonance
may contain four-quark admixture.

In addition, this large inclusive non-$D\bar D$ decay branching fraction may also indicate
that there are some new structure in addition to the conventionally dominated $1^3D_1$ state.
The huge branching fractions for $\psi(3770) \rightarrow {\rm VP}$ decays, 
extracted out from the CLEO measurement of the cross sections for these processes 
by Zhang's global amplitude analysis, also favour the assumption of that $\psi(3770)$
may contain four-quark admixture, or indicate that there may be some new structure near 3.770 GeV 
in addition to the conventionally dominated $1^3D_1$ state.

To search for more exclusive light hadron decay processes of the $\psi(3770)$,
one needs more data to be taken at both 3.773 and near 3.650 GeV at least. The best
way to search for more exclusive light hadron decay modes of $\psi(3770)$
is to make a fine energy scan over the energy range from 3.650 to 3.88 GeV
covering both the $\psi(3686)$ and $\psi(3770)$ resonances. This will be done
at the BES-III experiment in the near future.

\end{document}